%% file: main.tex
\documentclass[manuscript]{acmart}
\AtBeginDocument{%
  \providecommand\BibTeX{{%
    \normalfont B\kern-0.5em{\scshape i\kern-0.25em b}\kern-0.8em\TeX}}}

\usepackage{soul}

\newcommand{\pbyc}{\textit{PwR}}

\newcommand{\fontttc}[1]{\texttt{#1}}

\newcommand{\tcbtppink}[1]{\textcolor{magenta}{\st{#1}}}
\newcommand{\tcbtpgreen}[1]{\textcolor{teal}{\textbf{#1}}}

\begin{document}

\title{\pbyc{}: Exploring the Role of Representations in Conversational Programming}

\author{Pradyumna YM}
\author{Vinod Ganesan}
\author{Dinesh Kumar Arumugam}
\author{Meghna Gupta}
\author{Nischith Shadagopan}
\author{Tanay Dixit}
\author{Sameer Segal}
\email{sameersegal@microsoft.com}
\author{Pratyush Kumar}
\author{Mohit Jain}
\email{mohja@microsoft.com}
\author{Sriram Rajamani}
\email{sriram@microsoft.com}

\affiliation{%
  \institution{Microsoft Research}
  \city{Bangalore}
  \country{India}
}








\renewcommand{\shortauthors}{Pradyumna and Ganesan, et al.}

\begin{abstract}
  \input{content/0_abstract}
\end{abstract}

\ccsdesc[500]{Human-centered computing~Natural language interfaces}
\ccsdesc[300]{Human-centered computing~Empirical studies in HCI}
\ccsdesc[300]{Human-centered computing~Interactive systems and tools}

\keywords{LLM, Large Language Models, Programming, Natural Language Programming, Prompt Engineering, Novice Programmers, Chatbot Development, Human-AI Interaction, Design, Tool, Evaluation}


\maketitle

\newcommand{\sriram}[1]{\textcolor{red}{SR: \textit{#1}}}

\input{content/1_introduction}
\input{content/2_relatedwork}
\input{content/3_systemdesign}
\input{content/4_studydesign}
\input{content/5_findings}
\input{content/6_discussion}
\input{content/7_conclusion}
\begin{acks}
Thank you all our study participants for their time and patience.
\end{acks}

\bibliographystyle{ACM-Reference-Format}
\bibliography{references}


\end{document}

%% file: content/0_abstract.tex
Large Language Models (LLMs) have revolutionized programming and software engineering. AI programming assistants such as GitHub Copilot X enable conversational programming, narrowing the gap between human intent and code generation. However, prior literature has identified a key challenge--there is a gap between user's mental model of the system's understanding after a sequence of natural language utterances, and the AI system's actual understanding. To address this, we introduce Programming with Representations (\pbyc{}), an approach that uses representations to convey the system's understanding back to the user in natural language. We conducted an in-lab task-centered study with 14 users of varying programming proficiency and found that representations significantly improve understandability, and instilled a sense of agency among our participants. Expert programmers use them for verification, while intermediate programmers benefit from confirmation. Natural language-based development with LLMs, coupled with representations, promises to transform software development, making it more accessible and efficient.

%% file: content/1_introduction.tex
\section{Introduction}
Large Language Models (LLMs) such as Codex~\cite{codex}, GPT-4~\cite{gpt4}, and PaLM~\cite{palm} are disrupting programming and software engineering.
Powered by these LLMs, products such as GitHub Copilot X\footnote{GitHub Copilot X: Your AI pair programmer is leveling up: https://github.com/features/preview/copilot-x} and research tools like The Programmer's Assistant~\cite{programmer-assistant-iui23}
offer a conversational chatbot-like interface to programming, enabling the programmers to express their intent in a sequence of natural language utterances, initiating the generation or modification of code in response.
This paradigm shift in interaction has fundamentally transformed the programming experience, bridging the gap between human intent and code implementation~\cite{progwithAI-arxiv23, programmer-assistant-iui23}, ushering in a new era of natural language-based programming. This evolution holds the potential to democratize software development, catering to a broader audience, including those who possess domain expertise but lack extensive programming proficiency~\cite{democratize-prog-newstack}.

The recent emergence of these tools has piqued growing interest in the field of HCI and CSCW. 
Researchers have explored various facets of these AI-based programming assistants, including usability~\cite{usability-aiprog-icse24, usability-code-chiea22}, learnability~\cite{copilot-study-tochi23, progsynth-novice-uist22}, and trustworthiness~\cite{dev-trust-arxiv23, alignment-trust-liu2023}.
A key challenge that has been identified with these approaches is the gap between user's mental model of the system's understanding after a sequence of utterances, and the AI system's actual understanding~\cite{badpa-chi16, convey-chi18, johnny-cant-prompt-chi23, abstractiongap-chi23}.
As a result, users are often uncertain about the system's cognitive capabilities. 
Additionally, as existing LLM-based systems predominantly cater to intermediate and expert programmers, this gap particularly widens for novice programmers due to their limited understanding of the different software development constructs
\cite{johnny-cant-prompt-chi23, abstractiongap-chi23}.

Consider an example, where we interact with a conversational AI system to build a quiz game bot.
Suppose we start with the following sequence of two utterances: 
\begin{enumerate}
\small
\item[$U_0$:] \fontttc{Build a quiz game where you ask 20 questions about History.}
\item[$U_1$:] \fontttc{Give 10 points for each correct answer.}
\end{enumerate}
Now, suppose we want to make the game more challenging for users who answer questions correctly.
We can do this by adding two more utterances:
\begin{enumerate}
\small
\item [$U_2$:] \fontttc{Start questions with difficulty level 1.}
\item [$U_3$:] \fontttc{If the user gets two consecutive questions correct, increase the difficulty level of subsequent questions by 1.}
\end{enumerate}
Now, we realize that as we increase the difficulty level, we also want to increase the number of points for each correct answer.
However, we have already specified in $U_1$ that we want to give 10 points to each correct answer.
Consequently, we need to revise that utterance.
We are now unsure how to do this.
One way to do this is by using such an utterance:
\begin{enumerate}
\small
\item [$U_4$:] \fontttc{Instead of awarding 10 points for each correct answer, award points valued at 10 times the difficulty level.}
\end{enumerate}
However, with that utterance, the user is unsure whether the scoring rule has been correctly updated or not, and the user needs to test the bot by answering at least three questions to test the new scoring rule based on the difficulty level. 

\begin{figure}[h]
    \centering
    \includegraphics[width=0.6\linewidth]{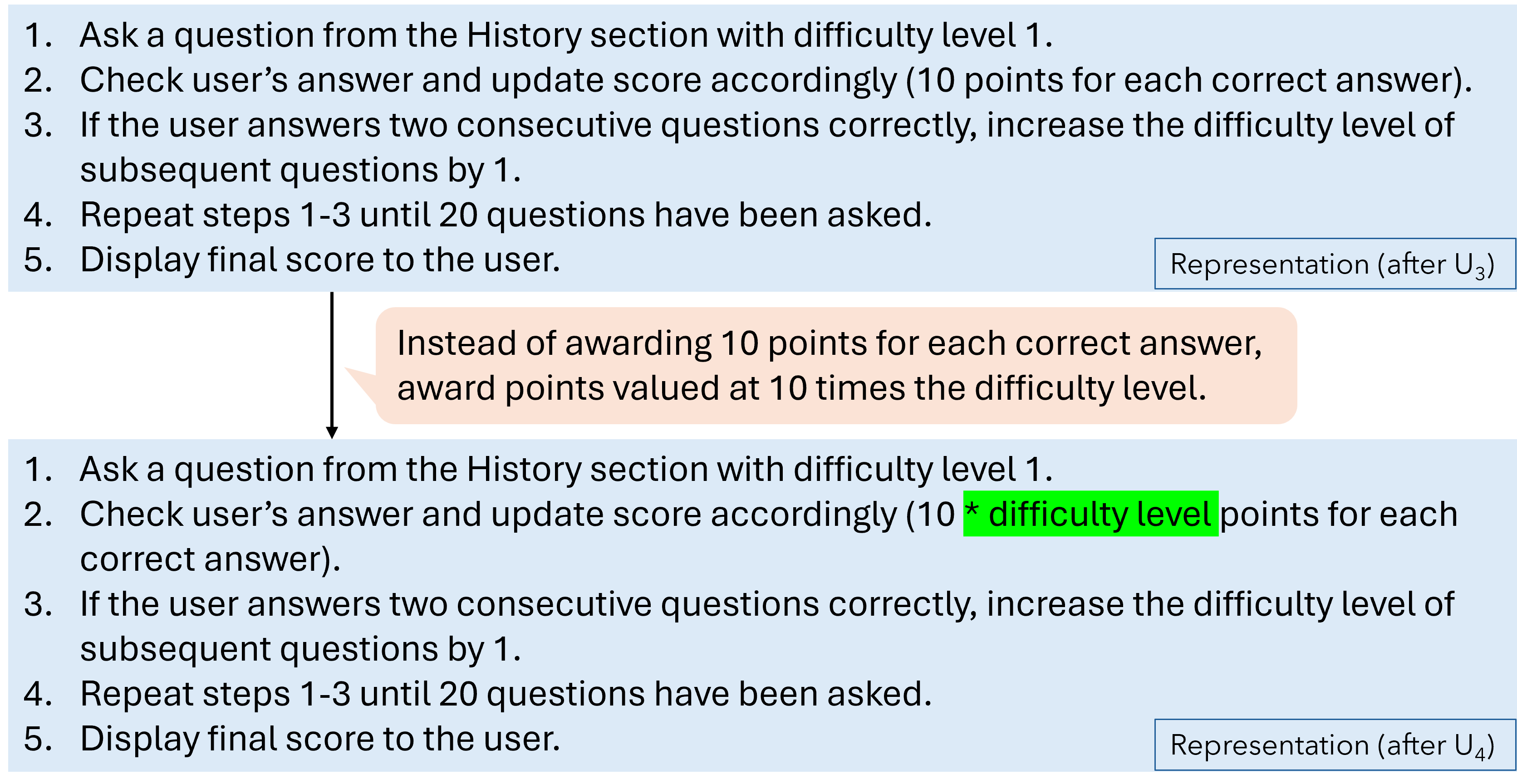}
    \caption{An example of \pbyc{} representation. A natural language utterance or direct edit can result in changes to the representations.}
    \label{fig:representation_edit}
\end{figure}

One way to bridge this gap is for the LLM-powered system to convey what it has understood from the conversation, such that the user and the AI system are ``on the same page'', \textit{i.e.}, share a mutual understanding
\cite{convey-chi18, abstractiongap-chi23}.
Expanding upon this concept, we propose a methodology called \pbyc{} (Programming with Representations, read as `\textit{power}'), where we use {\em representations} as a shared understanding between the user and the AI system, for facilitating conversational programming. 
A representation denotes the AI system's understanding of a sequence of utterances.
As shown in the top part of Figure~\ref{fig:representation_edit}, this logical representation is in natural language and captures the game details defined by the utterances $U_0$ to $U_3$.
With utterance $U_4$, the change in the representation, as shown in the bottom part of Figure~\ref{fig:representation_edit}, assures the user of the AI system's understanding.
Without such a representation, the only way for the user to gauge its understanding is to manually test the game and observe its behavior, which can be tedious and wasteful.
Alternatively, instead of using the utterance for editing, the user can edit the representation directly, replacing `\fontttc{10 points}' with `\fontttc{10 * difficulty level}' to achieve the same outcome.    
We believe that explicitly codifying the AI system's understanding as a (stylized) natural language representation could narrow the gap between the user's understanding of the system's capabilities and the system's actual capabilities.
This would further enable users to perform revisions and refine the representations (either by editing the representations directly, or indirectly through utterances) with the goal of completing their task.

To evaluate the usability of the \pbyc{} approach, we designed and implemented a \pbyc{} tool, enabling users to develop LLM-based chatbots through a combination of natural language utterances and representation edits.
We conducted an in-lab task-centered study with 14 participants having varying levels of programming expertise.
During the study, each participant created two chatbots--an interactive quiz bot and an informational knowledge bot--using the \pbyc{} tool and participated in a follow-up focus group discussion, delving into their experiences, workflows, challenges, and merits of using representations in chatbot development.
Our findings unveiled insights around the utilization of representations and uncovered distinct workflows employed by users. All our participants found representations helped increase their ``understandability'' of the chatbot development process.
This increased transparency instilled a sense of agency among them.
In addition, we observed a correlation between participants' level and style of representation usage with their programming expertise.
For instance, expert programmers read the representations to verify and debug the system's understanding, while intermediate and novice programmers quickly `glanced' over the representations, primarily to confirm if any portion of the representation had been modified.
With respect to edits, most participants preferred to edit the representations through natural language utterances, while only a few expert programmers decided to edit the representations directly to obtain higher efficiency from the \pbyc{} system.
Overall, natural language based software development using LLMs has potential to democratize the field, and representations offer an interesting sweet spot in the design space to bridge the gap between the LLM-powered system's interpretation of the natural language utterances and the user's understanding of it.



%% file: content/2_relatedwork.tex
\section{Related Work}
In this section, we present an overview of research related to natural language programming utilizing Large Language Models (LLMs), with a focus on chatbot development. We delve into the identified challenges and proposed solutions of this approach in the literature, to motivate and situate \pbyc{}.

\subsection{Natural Language Programming}
The endeavor to enable programming in natural languages has been a longstanding aspiration in computer science~\cite{ballard1979programming, lieberman2006feasibility}.
LLMs have significantly advanced this pursuit.
Codex~\cite{codex}, developed by OpenAI, powers GitHub Copilot, which enables programmers to write natural language comments to receive inline code suggestions~\cite{PengCopilot23}.
Several alternatives to Copilot, including Amazon CodeWhisperer, Codeium, and Tabnine, have been introduced.
Despite being a recent innovation, AI programming assistants have garnered substantial popularity with more than a million developers\footnote{Note: We use the term `programmers' and `developers' interchangeably.} using GitHub Copilot~\cite{TheEconomicGithubCopilot}.
Studies have demonstrated a significant increase in developers' productivity when using AI programming assistants, with tasks being completed up to 55.8\% faster, 28.8\% of GitHub Copilot's Python suggestions being accepted, and an average of 46\% of developers' code being generated using Copilot~\cite{PengCopilot23, productivity_sigplan22}.
In addition, LLMs have catalyzed the emergence of numerous low-code and no-code frameworks.
Commercial products such as GitHub Copilot X and Microsoft's Power Apps AI Copilot\footnote{Power Apps - AI Copilot overview: https://learn.microsoft.com/en-us/power-apps/maker/canvas-apps/ai-overview}
leverage LLMs to facilitate the development of diverse web and mobile applications using conversation with a chatbot in natural language referred to as \textit{conversational programming}.
Even recent research efforts explore similar directions, such as 
The Programmer's Assistant~\cite{programmer-assistant-iui23} for programming tasks.
These tools empower individuals, even those with no prior programming experience, to develop applications, spanning from web applications to chatbots, simply by conversing with a chatbot.

With that, HCI and CSCW researchers have explored various aspects of natural language to code generation tools, including usability~\cite{usability-aiprog-icse24, usability-code-chiea22, explainability-iui22}, learnability~\cite{copilot-study-tochi23, progsynth-novice-uist22}, and trustworthiness~\cite{dev-trust-arxiv23, alignment-trust-liu2023}.
This extensive research is integral in ensuring that these tools not only perform effectively but are also user-friendly and reliable.
\citet{usability-aiprog-icse24} surveyed 410 developers and \citet{usability-code-chiea22} interviewed 24 developers to understand the usability of these tools. 
Their findings revealed that developers use AI-based code generation tools to reduce keystrokes, expedite programming tasks, recall syntax, and obtain a valuable starting point, thus saving time otherwise spent searching online.
However, they also pointed out that developers faced challenges with these tools, 
mainly difficulties in understanding, editing, and debugging generated code snippets, and struggles in controlling the tool to produce the desired output.

In addition to usability, establishing and maintaining an appropriate level of trust in AI programming assistants is crucial. Inadequate trust can deter developers from leveraging AI to increase productivity, while excessive trust may lead to overlooking potential risks and security vulnerabilities.
Based on interviews with 17 developers, \citet{dev-trust-arxiv23} found that developers rely on shared experiences within the community and community feedback to evaluate AI tools. 
Trust also plays a fundamental role in the learnability of AI programming assistant tools~\cite{progsynth-novice-uist22}. For instance, these tools have also been lauded for their potential to reduce entry barriers into programming, thereby enhancing accessibility for a wider audience.
\citet{progsynth-novice-uist22} asked novice programmers to interact with five such tools, followed by interviews to assess their learnability.
They discovered that novices often struggle with formulating natural language prompts, especially in determining which types of prompts are effective.
(Note: \textit{Prompts} are inputs in natural language that users can provide to elicit specific responses from an LLM.)
They also noted that using such tools requires `\textit{more effort and }(results in) \textit{less trust}'.
AI program assistants require users to provide specification/prompt, which adds an extra layer of work for the user.
While the generated code may fulfill the given specification, it may not necessarily align with the user's original intent~\cite{alignment-trust-liu2023}.
We will delve deeper into a similar issue in the following section.


\subsection{Identified Gaps and Proposed Solutions}
In the pre-LLM era of chatbots, researchers noted a significant gap between user experience and user expectation with respect to the understanding of chatbot~\cite{convey-chi18, allworknoplay-chi18, badpa-chi16, informing-chatbot-design-dis18}.
There exists a mismatch between the chatbot's understanding and the user's perception of the chatbot's understanding.
As a result, users often found it challenging to employ chatbots for complex tasks, due to uncertainty about the chatbot's capabilities.
Studies focusing on the usability, trustworthiness, and learnability of LLMs discerned a similar gap~\cite{dev-trust-arxiv23, usability-aiprog-icse24, usability-code-chiea22}.
After providing a prompt, users have been found to remain uncertain about whether the LLM-powered system accurately comprehended it~\cite{johnny-cant-prompt-chi23, abstractiongap-chi23}.
In addition, while prompting LLMs may seem straightforward, formulating effective prompts has proven to be a challenging task~\cite{johnny-cant-prompt-chi23}.
Even when users have a clearly-defined intent, choosing a prompt from the vast array of possible natural language utterances that they believe the system will reliably interpret to yield a satisfactory outcome was found to be daunting~\cite{abstractiongap-chi23}.

To address these issues, \citet{abstractiongap-chi23} introduced the concept of `\textit{grounded abstract matching}',
to translate the user's utterance into a systematic and predictable natural language utterance based on the LLM's understanding.
This serves as an editable illustration of how to consistently trigger the same action.
Through a user study with 24 participants interacting with LLMs in natural language to generate spreadsheet formulas, they found that this grounded approach improved users' ability to recover from system failures and that users experienced greater confidence and a heightened sense of control when using the system.
While being inspired by similar considerations, \pbyc{}'s task of developing a chatbot application
is more complex than a single-turn conversation to generate spreadsheet formulas.

Addressing the issue of aiding novice programmers in devising effective prompt strategies, \citet{johnny-cant-prompt-chi23} proposed BotDesigner, a no-code LLM-based tool that enables users to develop a chatbot through prompts alone. 
Using this tool, they observed 10 participants lacking prior experience in prompt design, execute a chatbot design task. Their findings suggest that while end-users can experiment with prompt designs opportunistically, they encounter challenges in making robust and systematic progress.
In a similar vein, drawing inspiration from the grounded abstraction matching framework~\cite{abstractiongap-chi23}, our proposed system \pbyc{} attempts to bridge the gap between developer's perception of the LLM's understanding and the LLM's actual understanding, in the process of facilitating chatbot development through natural language interaction (similar to \cite{johnny-cant-prompt-chi23}).
\pbyc{} entails displaying the current comprehension of the AI system to the user in a structured natural language format, referred to as `representations'.
Additionally, it empowers users to edit the representations either directly, or indirectly, using conversations.

%% file: content/3_systemdesign.tex
\section{System Design}
To streamline the process of developing chatbots through natural language interaction with a chatbot and to provide a visual representation of the generated chatbot's capabilities and mental model, we created the \pbyc{} tool.
The representation within the tool consists of three components (as shown in the middle `Representation' pane in Figure~\ref{fig:pbyctool}):
\begin{itemize}
    \item A knowledge base [KB], consisting of a set of key-value pairs. The keys denote distinct section names, and the corresponding values are strings containing knowledge about that section, articulated in natural language.
    \item The logic [Logic] of the bot, encompassing a set of rules expressed in natural language. These rules delineate the set of possible behaviors that the chatbot can exhibit.
    \item A set of variables [Variables], which stores the conversational state. Each variable has its initialization and update rules written in natural language, providing guidelines for how its value should be established and modified over the course of the conversation.
\end{itemize}
These three components were chosen for representations as they align with a conventional coding structure, comprising of variables, business logic, and data (knowledge base), thereby facilitating a familiar and structured approach to chatbot development.
Below, we provide an overview of the tool's user interface and its implementation details.

\subsection{User Interface}
The primary objective of the \pbyc{} tool is to furnish an interactive environment for chatbot development and testing. 
The tool comprises of two main pages: the template selection page and the \pbyc{} chatbot builder page.

\textbf{Template Selection Page}:
Upon signing up and logging into the \pbyc{} tool, users are directed to the landing page, which presents a catalog of available templates.
Additionally, it displays a list of projects initiated by the user.
Presently, the tool offers various templates, such as quiz bot, e-commerce bot, and knowledge bot.
Each template has an initial (hidden) prompt, designed to streamline the process of constructing customized bots.
For instance, a part of the prompt text for the quiz bot template is as follows: `\fontttc{You are a bot designed to develop a chat-based game defined by a knowledge base [KB] and a set of logical rules [Logic]... [Logic] governs the behavior of the bot... The logic uses variables defined in [Variables]... We want to ensure that the set of rules do not contradict each other.}'
Users can initiate a \pbyc{} project by selecting a template of their choice and specifying a project name, with an optional brief description.
This landing page allows users to seamlessly resume their work on various projects across multiple sessions.

\begin{figure}[!tbp]
    \centering
    \includegraphics[width=0.9\linewidth]{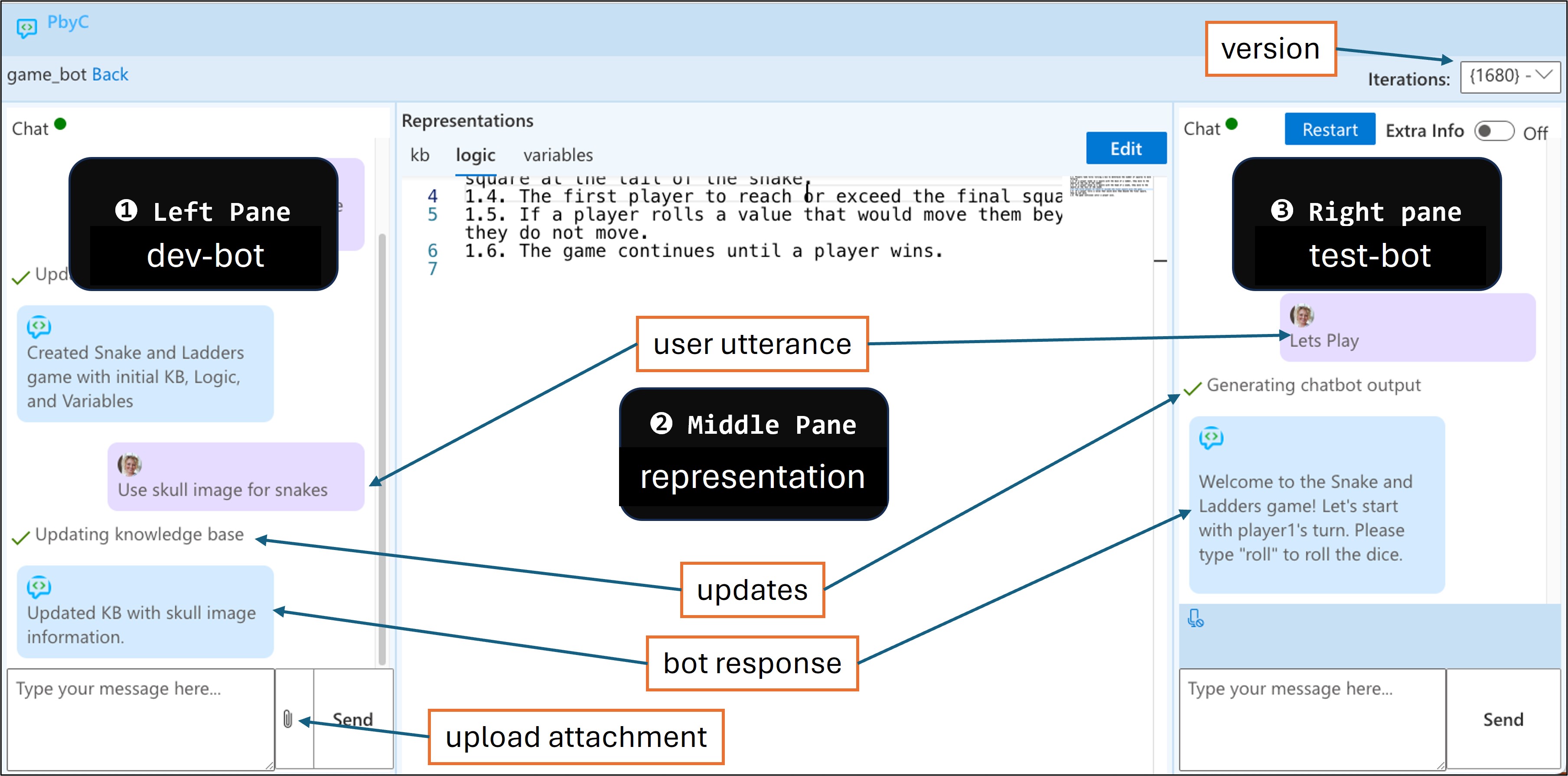}
    \caption{A screenshot of the \pbyc{} tool, chatbot builder page.}
    \label{fig:pbyctool}
\end{figure}

\textbf{Chatbot Builder Page}:
The chatbot builder page consists of three vertical panes (Figure~\ref{fig:pbyctool}): the \textit{dev-bot} on the left, the \textit{representations} in the center, and the \textit{test-bot} on the right.
\begin{itemize}
    \item The dev-bot is a chat interface for the user to input utterances specifying rules/instructions for chatbot development, and for the system to provide responses to these inputs. Each utterance to the dev-bot triggers updates to the representations.
    \item The middle pane is designated for viewing and editing the three components--the knowledge base (KB), logic, and variables--of the representations.
    \item The test-bot is the functional output bot, created using the user's utterances to the dev-bot and any direct representation edits. It strictly adheres to the rules specified in the representation components present in the middle pane. This feature enables the user to assess the current capabilities of the output bot, primarily for testing and debugging purposes. Each utterance directed to the test-bot results in the execution of representations.
\end{itemize}

We anticipated that $\sim$5 users would concurrently use \pbyc{} during the study, thereby sharing the same OpenAI key. 
Given that each utterance to the dev-bot and every representation edit trigger multiple calls to GPT-4, we foresaw potential delays in receiving the system's response.
To address this, we incorporated \textit{updates} in both the dev-bot and test-bot views (Figure~\ref{fig:pbyctool}), providing progress notifications for each GPT-4 call.
It is similar to the `update' feature employed by Microsoft's Bing Chat, which keeps users informed about the progress of their queries.

The \textit{response} from the dev-bot combines the dev-bot's comprehension of the user's utterance with a summary of the changes made in the representations.
On the other hand, the response from the test-bot is contingent on the output bot's capabilities as reflected in the representations.
The logic component is the default view in the representation, and only one component can be edited at a time to minimize inconsistencies between components.
An edit to any component in the representation may lead to corresponding updates in other component(s), and in some cases, even all components.
Note: The \pbyc{} system uses calls to LLM to enforce the invariant that the logical rules within the representation components do not have any inconsistency.
The dev-bot also offers the capability to upload attachments, which serve as a KB.
This feature proves particularly valuable in the context of developing a custom knowledge base.

In the top right corner, two interesting features are available.
The \textit{Version} drop-down allows users to access different versions of the output bot, each corresponding to a specific state of the representation after an utterance to the dev-bot or direct representation edit.
This enables users to revert to a previous state of the output bot at any given time.
The \textit{Restart} button facilitates a user in resetting their conversation with the test-bot.
It deletes the test-bot's current context and reloads the representation, providing a clean slate for retesting the output bot.

\subsection{Implementation Details}
The \pbyc{} tool was developed as a web application utilizing React, an open-source front-end JavaScript library.
The server-side back-end was implemented using Python.
OpenAI's GPT-4 Large Language Model (LLM) played a pivotal role, with each utterance to the dev-bot and each direct representation edit leading to multiple calls to GPT-4.
All input and output events, including user utterances to both dev-bot and test-bot, bot responses, representation states and edits, and user clicks, were meticulously logged and stored on the server for subsequent analysis.

%% file: content/4_studydesign.tex
\section{Study Design}

\subsection{Procedure}
To gain insights into how end users use \pbyc{} and approach representations, we recruited 14 participants possessing varied degrees of experience in programming.
To recruit a diverse set of participants, we emailed individuals across diverse roles within our organization, including designers, engineers, researchers, and operations, HR, and finance staff.
We received responses from 18 individuals, out of which 14 participated in the study.
Four individuals were not able to participate due to scheduling constraints.

We conducted an in-lab task-centered investigation comprising of three user study sessions.
The initial session incorporated 4 participants, whereas both session-2 and session-3 accommodated 5 participants each.
For each session, a subgroup of participants was requested to convene in a conference room (with a seating capacity of 12 individuals) with their laptops.
These subgroups were formed by considering a mix of programming proficiency levels, ensuring that each subgroup consisted of one or two participants identified as novice, intermediate, and expert programmers.
Participants were provided with notebooks and pens for note-taking.
Throughout the study, two or more researchers were physically present in the room to help participants with their queries, document their observations, and conduct focus group discussions.
We selected the task of creating two chatbots: (1) an interactive quiz bot, and (2) an informational knowledge bot.
These tasks were curated to encompass a majority of \pbyc{} features, enabling open-ended exploration of its capabilities.
Additionally, to cater to various programming skill levels, the tasks were designed to be sufficiently approachable for novice and intermediate programmers, while also allowing expert programmers to introduce intricacies.

With verbal agreement from the participants, we initiated a Microsoft Teams call in the conference room, and activated the `Start recording' feature.
After signing the online consent form, the session commenced with a 2-minute instructional video, showcasing key \pbyc{} features through an example of a tic-tac-toe bot development.
Participants were encouraged to seek clarifications during the video, and the video was paused to address queries.
Afterward, participants accessed \pbyc{} using their corporate login credentials.
For the first task, a 1-minute demonstration video of a `Who Wants to Be a Millionaire?' quiz bot developed with \pbyc{} was presented. 
The video was shown to mitigate the `Blank Page Syndrome'~\cite{blank-page-syndrome}, wherein users confront difficulty in commencing a task.
Subsequently, participants had 5 minutes to conceptualize and outline rules (including game play, lifelines, and scoring) for their personalized quiz bots.
After that, participants were tasked to create that quiz bot within a 30-minute timeframe using \pbyc{}.
The second task involved developing a knowledge bot that utilizes a custom knowledge base to respond to queries, diverging from the generic knowledge of the LLM.
This is a common use case of chatbots in a variety of domains including medicine, legal, and customer support.
Participants spent 5 minutes devising scenarios for their knowledge bots, including themes, applicable knowledge bases, and supported question types.
Following this, participants were allocated 30 minutes to actualize their knowledge bots using \pbyc{}.
Each task segment was punctuated by a 5-minute break.

Upon completing both tasks, participants filled out an online demographic questionnaire and a NASA-TLX 5-point Likert-scale ratings questionnaire~\cite{nasa-tlx}, which took $\sim$10 minutes.
The study concluded with a 60-minute focus-group discussion, delving into participants' holistic experiences, challenges faced while acquainting themselves with \pbyc{}, their usage of representations, and their workflow for bot development. 
The focus group discussions were transcribed, soon after they were
conducted.
The study was conducted in English.
Participants were not paid for participation.

\input{tables/participants}
\subsection{Participants}
Fourteen individuals (7 female and 7 male) participated in the study.
Among our participants, 3 self-identified as novice programmers, 6 as intermediate programmers, and 5 as expert programmers.
With respect to prior experience in prompt engineering, 5 of them had no experience, 6 had moderate experience interacting with Microsoft Bing and OpenAI ChatGPT playground, and 3 had extensive experience using it on a daily basis for their research projects.
For more details on the demographic composition of our participants, please refer Table~\ref{table:participants}.

\subsection{Data Analysis}
We conducted a mixed-method analysis to systematically analyze the collected data. The logs generated by \pbyc{} tool interactions were quantitatively analyzed.
We used descriptive statistics and statistical tests (like t-tests and ANOVAs) on the number of messages sent to dev-bot and test-bot, number of words exchanged, and direct edits to representations.
For qualitative analysis, we subjected our focus group discussion transcription data and researcher's observation notes to open coding, and rigorously categorized our codes to understand user behaviour. Two authors participated in the coding process and iterated upon the codes until a consensus was reached. Over the course of analysis, they discussed coding plans, developed a preliminary codebook, reviewed the codebook, refined/edited codes, and finalized categories and themes. The first-level codes were very specific, such as ``democratize programming'', ``glancing representations'', and ``interaction with dev-bot''. After several rounds of iteration, the codes were condensed into high-level themes, such as ``representation usage and its limitations'', ``bot development workflows'', and ``programming experience with \pbyc{}''.

%% file: tables/participants.tex

\begin{table}[]
\centering
\caption{Participants demography, along with the type/theme of quiz bot and knowledge bot they created. {[}* CHW: Community Health Worker, HCI: Human-Computer Interaction, TA: Teaching Assistant, PVI: People with Vision Impairments, NLP: Natural Language Processing, IT: Information Technology{]}}
\label{table:participants}
\resizebox{\textwidth}{!}{%
\begin{tabular}{ccclllc|ll}
\hline
\textbf{PId} &
  \textbf{Sex} &
  \textbf{Age} &
  \textbf{Education} &
  \textbf{\begin{tabular}[c]{@{}l@{}}Programming\\ Level\end{tabular}} &
  \textbf{\begin{tabular}[c]{@{}l@{}}Prompting\\ Experience\end{tabular}} &
  \textbf{\begin{tabular}[c]{@{}c@{}}Session\\ \#\end{tabular}} &
  \textbf{Quiz Bot} &
  \textbf{Knowledge Bot} \\ \hline
P1  & F & 18-25 & Bachelors & Novice       & No exp       & S1 & Bollywood songs         & Lesson planner          \\
P2  & F & 18-25 & Bachelors & Intermediate & Extensive       & S1 & Indian history, Friends & CHW* trainer            \\
P3  & M & 18-25 & Masters   & Intermediate & Moderate & S1 & Mentalist screenplay    & Reddit thread explainer \\
P4  & M & 18-25 & Bachelors & Expert       & Extensive       & S1 & Math/logic puzzles      & Bhagavad Gita           \\
P5  & F & 18-25 & Bachelors & Novice       & Moderate & S2 & Mahabharata             & HCI* course TA*         \\
P6  & M & 18-25 & Bachelors & Intermediate & Moderate & S2 & Football                & Finance advisor         \\
P7  & F & 18-25 & Bachelors & Intermediate & No exp       & S2 & Animal                  & Maths teacher for PVI*  \\
P8  & M & 18-25 & Bachelors & Expert       & No exp       & S2 & Bollywood               & Lesson planner          \\
P9  & M & 18-25 & Bachelors & Expert       & Moderate & S2 & Punjabi language        & NLP* teacher            \\
P10 & F & 36-45 & Masters   & Novice       & No exp       & S3 & Cricket                 & IT* help                \\
P11 & F & 18-25 & Masters   & Intermediate & Moderate & S3 & Geography               & Programming TA*         \\
P12 & F & 18-25 & Bachelors & Intermediate & Extensive       & S3 & Indian folk dance       & Dance exam prep         \\
P13 & M & 18-25 & Bachelors & Expert       & No exp       & S3 & Student welfare         & Github project          \\
P14 & M & 18-25 & Bachelors & Expert       & Moderate & S3 & Random                  & Science fiction         \\ \hline
\end{tabular}%
}
\end{table}

%% file: content/5_findings.tex
\input{tables/results}
\section{Findings}
Overall, our participants created 33 chatbots (17 quiz bots and 16 knowledge bots), sent 214 messages to the dev-bot (average length = 21.1$\pm$16.0 words/message) and 424 messages to the test-bot (average length = 3.7$\pm$4.5 words/message), and edited representations directly 42 times.
Please refer to Table~\ref{table:results} for more details.
Due to the flexibility offered in the study, our participants developed a variety of quiz bots and knowledge bots, ranging from mythology (Mahabharata quiz bot, Bhagavad Gita knowledge bot) to education (Punjabi language quiz bot, HCI TA knowledge bot). Information regarding the themes of the bots developed by each participant can be found in Table~\ref{table:participants}.
Below, we discuss findings related to our participants' workflows in building these bots, the role of representation and its usage, and the impact of programming proficiency on participant's \pbyc{} experience.
(Note: We use the \fontttc{Typewrite font} for user's utterances and \pbyc{}'s representations. Additionally, edits/updates to text are indicated using \fontttc{\tcbtpgreen{green with bold}} for added text and \fontttc{\tcbtppink{pink with strikethrough}} for deleted text.)

\begin{figure}[!tbp]
    \centering
    \includegraphics[width=1\linewidth]{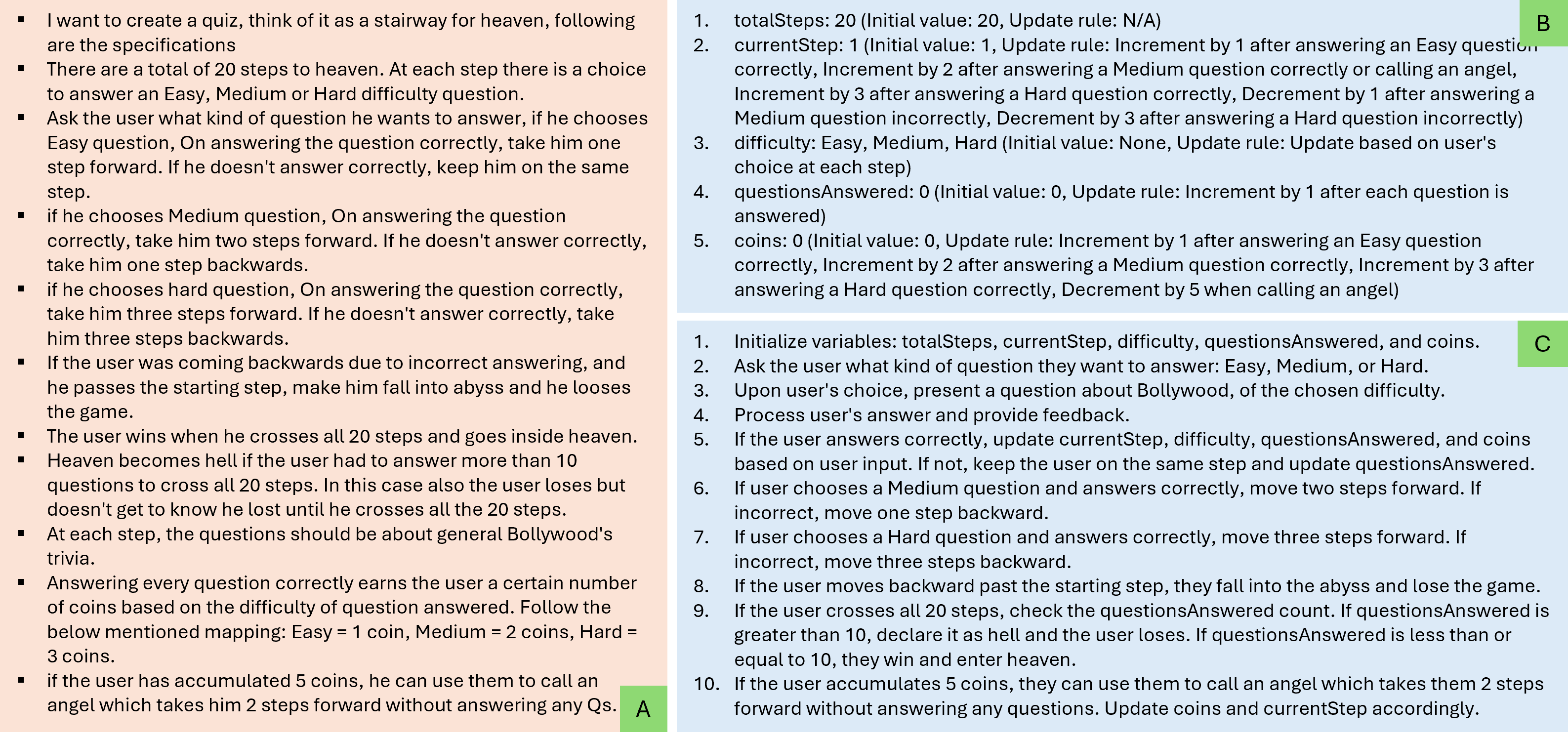}
    \caption{Example of a quiz bot created by P8: (A) Utterances to the dev-bot. Representation views: (B) Variables, and (C) Logic.}
    \label{fig:sample_rep}
\end{figure}

\subsection{Representations Usage and Limitations}
\label{sec:rep_usage}
A majority of our participants highly appreciated the representations and used them in novel ways.
The three key reasons behind using representations were: (1) to gain insights into the system's understanding, (2) to identify and address instances where the communication with the dev-bot failed, and (3) to obtain higher efficiency from the \pbyc{} tool. We discuss them in detail below.

\subsubsection{Increased Understanding}
\label{section:understanding}
A majority of our participants (12 out of 14) were of the opinion that representations helped in increasing their ``\textit{understandability}'' of the chatbot development process and the system's internal interpretation and logic.
This increased transparency instilled a sense of agency among our participants and provided them a greater control over the development process.
This is in contrast with prior findings on prompt engineering with LLMs which identified a lack of control among its users~\cite{hbr-prompt-notfuture, usability-aiprog-icse24, explainability-iui22, johnny-cant-prompt-chi23}.
In particular, programmers have complained about how prompting is opaque and diminishes their control.
Representations not only helped our participants to gain insights into the system's understanding but also equipped them to modify that understanding directly, obviating multiple rounds of interactions with the dev-dot and test-bot.
P11 echoed that sentiment:
\begin{quote}
    ``It just gave me some sense of control, that it's not a complete black box. It's not just like on the left [dev-bot] I'm giving some instructions and by `magic' it's giving me something to use on the right [test-bot]... As P14 said, we are programmers by nature, so our first instinct is to look at the logic, the variables, and confirm that whatever we give in that natural language, it's interpreting correctly.'' (P11)
\end{quote}

Similar to P11, most of our participants recalled ``\textit{glancing}'' at the generated representations to validate the understanding of \pbyc{}'s system.
However, the level of glancing varied across our participants.
We found a correlation between the level of glancing and their programming expertise, with expert programmers reading the representations to verify and debug the system's understanding, intermediate programmers peeking into representations just to check if ``\textit{something got edited}'' after each utterance to the dev-bot, and novice programmers (almost) ignoring it. For instance:
\begin{quote}
    ``I gained a lot from these representations, especially the logic tab. It really helped in understanding whatever I'm asking [the dev-bot]... if that's being processed or not correctly.'' (P9, Expert)
    
    ``I didn't read the KB, the logic, or the variables tab. But they made sense to me as I have some coding experience... When I prompted a query (messaged the dev-bot), I would just glance that something got edited there [in representations]. I would just switch the tabs and see that some line got added there which looks a little like what I had added. I did not really verify what's being added.'' (P2, Intermediate)
    
    ``I didn't even look at this part (representations) because I don't understand it. If it (dev-bot) said something got edited there, I was like good!'' (P1, Novice)
\end{quote}



One of the reasons mentioned by our participants for only glancing at the generated representations instead of reading them in entirety was that the content appeared to be verbose and ``\textit{very formal}'', making it cumbersome to read.
This formality and verbosity in the representations were intentionally designed to encompass and convey the user's intended meaning comprehensively.
To make the representations more consumable, participants suggested novel solutions, including adding a flow diagram as a separate representation to show the test-bot's dialog flow, and visualizing a ``\textit{diff}'' (wherein added text gets highlighted in green and removed text gets highlighted in red) in each of the representation tabs after each update.
P5 elaborated on it:
\begin{quote}
    ``After providing some additional instructions [to the dev-bot]... I can't see what specific changes were made [in representations]. I can see it in the [dev-bot output] message, but not in the [representation] tabs. So I think having a diff over it would make it easier to understand what actually happened.'' (P5)
\end{quote}


\subsubsection{Help with Debugging}
Apart from glancing at representations for increased understanding, participants mentioned taking a deep dive into it when the test-bot failed to work as per their expectations.
This deeper examination primarily aimed at pinpointing the underlying cause of the identified bug.
For instance, P2 mentioned that while interacting with the dev-bot, she would glance at the representations to ensure that ``\textit{something got edited}'', however while testing the generated quiz bot, she realized that the test-bot was not transitioning to the next question automatically:
\begin{quote}
    ``In my case, the score and the reasoning behind the correct answer was getting displayed after each question. However, it didn't itself move to the next question, even though I remember adding the automatic transitioning to the next question [logic] once it has displayed the score... I checked the logic tab and I realized I hadn't added it... I was like `oh shit', and then I added it.'' (P2)
\end{quote}
Here, there was a gap between the P2's perceived understanding of the bot's capabilities and the bot's actual capabilities.
\pbyc{} representations played a pivotal role in bridging that gap, thereby facilitating the process of debugging and rectifying the identified issue.
Prior work reported that expert programmers excel in debugging primarily due to their superior ability to comprehend the program~\cite{debugging-expert-novice-schi86}.
This discrepancy in P2's perception could also be attributed to her workflow, wherein she provided 15 back-to-back utterances to the dev-bot for constructing an Indian history quiz bot, which was then followed by an interaction with the test-bot.
Managing and retaining 15 instructions within memory can be challenging for a human, hence the representations served as a means to revisit and review these instructions.

Similarly, three participants reported that the intricate scoring scheme they had devised for their quiz bots was not being accurately reflected in the representation. Consequently, they took corrective actions by either engaging in a conversation with the dev-bot or directly modifying the logic within the representation. For instance:
\begin{quote}
    ``After giving a prompt [to dev-bot] to come up with a scoring system the way I wanted it to be, twice, I was unsure if it was understanding what I was asking it to do, and that's what made me go to the representation and check... then identify the incorrect representation and fix it... I was unsure if I was able to get my point across and hence I just crosschecked it in the logic [tab].'' (P6)
\end{quote}
The intricate logic was a key factor motivating our participants to actively engage in the process of (re-)reading representations and directly making edits to them.
For instance, P13, in the process of developing a knowledge bot to assist prospective students with information about his school, realized: ``\textit{At some point the logic got a bit too complicated... Like my instructions }(utterances) [to the dev-bot]\textit{ were overlapping, then I just edited the logic manually.}''

We observed instances wherein a new utterance to the dev-bot by our participants contradicted one or more of their previous utterances.
In such scenarios, the \pbyc{} system updated one or more components of representation to ensure coherence and consistency, thus enforcing the defined invariant.
For instance, P2's second utterance was: ``\fontttc{The theme of the quiz bot should be Indian history.}'', and later, on her 17th utterance, she requested, ``\fontttc{Make it a friends sitcom quiz}''.
This resulted in both the knowledge base---`\fontttc{Quiz Bot: A quiz bot with 10 questions in total, themed around \tcbtppink{Indian history}\tcbtpgreen{Friends sitcom}.}'---and logic---`\fontttc{The quiz bot should interact like a quizmaster, challenging the player with questions about \tcbtppink{Indian history}\tcbtpgreen{Friends sitcom}.}'---getting updated consistently.
In rare instances, inconsistency resulted in the representations getting overwritten to reach a consistent state.
For instance, P7 performed this edit in the \fontttc{logic} tab of representation:
\fontttc{`9. Award 30 coins for an answer. 10. Display an animal created out of symbols with the text `Congratulations' if the player answers 3 questions correctly in a row. \tcbtpgreen{11. Award 10 coins for every correct answer and a message saying `Fantastic Job!'}'}.
Here, logic-11 contradicts logic-9, resulting in the \pbyc{} system deleting both logic statements following this edit.
Instead of fixing the bot from that state, P9 reverted back to a previous working version of the bot's state ``\textit{to get all my logic and variables back}''.
Supporting rollbacks to a previous version of the generated bot has been made possible due to the well-defined representations for the bot.

\subsubsection{Faster Edits}
Three expert programmers mentioned that they chose to directly edit representations due to their perception of it being a ``\textit{faster}'' approach, both in terms of the time required to formulate the edit and the time to execute that edit.
P14 mentioned that for certain modifications such as altering the scoring scheme in the quiz bot (e.g., changing the point allocation from 10 to 100 for each correct answer), directly editing the representation was significantly faster than crafting an utterance for the dev-bot.
Furthermore, these expert programmers correctly deduced that utterances made to the dev-bot were subsequently converted into representations, which were then propagated to the test-bot.
Consequently, by directly modifying the representations, they effectively halved the time required for their edits to take effect.
Note: During the study, with 4-5 participants sharing the same OpenAI key and each utterance resulting in multiple calls to GPT-4, delays were experienced 
by our participants.
This may have also contributed to their preference for using representation edits as a workaround.

Despite the benefits and the rationale for directly editing representations, the majority of our participants exhibited a reluctance to do so.
In fact, two of our participants refrained from attempting any edits on representations altogether, and six participants engaged in only two or fewer representation edits.
A primary reason cited for this hesitancy was the perceived formality of the representations, which instilled concerns that their edits might inadvertently disrupt the bot's functionality.
Consequently, participants generally considered deletion and minor edits to be safer options in comparison to adding new content directly to the representations, as P5 described:
\begin{quote}
    ``I preferred not to add stuff on my own to the logic or the variables because I felt like it was writing things in a very formal way. So either I would just write it in the [dev] chat and let it (representations) do whatever it's doing, or max I will delete a few logic or do some very minor edits, but certainly not adding things directly to logic.'' (P5)
\end{quote}

\subsection{Chatbot Development Workflows}
Based on the quantitative analysis of the interaction log data, we identified four distinct workflows adopted by our participants.
To enhance the validity of these findings, we supplemented them with insights drawn from the focus group discussion data.
Notably, none of the participants confined themselves to a single workflow; instead, they explored various workflows.
Before delving into the description of the workflows, it is worth highlighting an interesting observation: participants consistently spent considerable time in each pane before transitioning to the next.
On analyzing consecutive interactions, we found certain interaction sequences to occur significantly more frequently (p<0.05) than others: a dev-bot utterance was directly followed by another dev-bot utterance 107 times, a representation click was succeeded by another representation click 144 times, and a test-bot utterance was immediately followed by another test-bot utterance 305 times.
These sequences statistically outweighed other consecutive interactions, such as dev-bot $\rightarrow$ representation (32 times), dev-bot $\rightarrow$ test-bot (68 times), representation $\rightarrow$ dev-bot (21 times), representation $\rightarrow$ test-bot (34 times), test-bot $\rightarrow$ dev-bot (82 times), and test-bot $\rightarrow$ representation (25 times).

\subsubsection{${(dev - rep_{glance})}^n$ -- test}
As mentioned earlier, our participants predominantly utilized the representation to gain a deeper understanding of how the system interpreted their utterances to dev-bot.
In terms of workflow, participants concentrated their interaction with the dev-bot, and regularly reviewed the representation edits resulting from their dev-bot utterances, as P5 stated: ``\textit{I preferred to manually check the [representation] tabs after each message [to dev-bot]... just as a confirmation, that OK, whatever I said has been taken into account}''.
This process was repeated iteratively.
Once they were satisfied with a version of the representation, they proceeded to test it by interacting with the test-bot.
(Note: Since our log data did not capture participant's glancing over the different representation components, unless they actively clicked on a component tab to switch, hence we do not report quantitative data regarding glancing behavior.)

Quantitatively, the logic component tab was clicked 92 times, followed by the variables tab (59) and the knowledge base (KB) tab (54).
On conducting a one-way ANOVA, we observed a significant effect of representation type on the number of clicks (F(2,27)=6.1, p<0.05). Note: Four participants did not click any of the representation tabs, however since the default view was set to the logic tab, they may have still glanced at or read the logic.
A Tukey's pairwise comparison revealed significant differences between the number of clicks on the logic tab and the variables tab (p < 0.05), as well as between the logic tab and the KB tab (p < 0.05).
Despite participants predominantly clicked on the logic tab, we found instances where participants explored the variables and KB tab for specific use cases. For example: 
\begin{quote}
    ``I wanted to create a [knowledge] bot to explain the tax code of India... It would have a lot of rules and sub-rules and exceptions, right? So I would not know what knowledge it has encoded because it is very huge... but I can learn by looking through it (KB tab).'' (P13)
\end{quote}

The act of glancing at the representations played a guiding role in our participants' actions.
This was not only limited to assessing readiness for interaction with the test-bot, but it also influenced their subsequent utterances to the dev-bot.
For instance, ``\textit{When I gave the instructions} [to dev-bot]\textit{, it formulated some logic... initially the logic was very broad, it wasn't actually doing things that I wanted it to... So consequently I provided more specific instructions} [to dev-bot].'' (P9).


\subsubsection{$dev^n$ -- $rep_{glance}$ -- test}
Three participants (P1, P2, and P8) with varying levels of programming expertise reported that their primary approach involved engaging in conversations with the dev-bot.
After several rounds of conversation, they glanced at the representation and proceeded to test the generated bot.
They largely disregarded the representation components in that phase, as they were deeply immersed in the bot-building process and preferred not to be interrupted, even by receiving implicit feedback from the captured representations.
For instance, P2 sent 15 utterances and P8 sent 10 utterances to the dev-bot before proceeding to test the output bot.
\begin{quote}
    ``For the quiz bot, I wasn't looking at the representations at all while writing the prompts... I was coming up with the rules and writing the prompts (utterances to dev-bot). I just wanted to complete my thought process first of all in the left-hand side and then go to see the logic (representations)... When I went there I was blown away... It (the representations) kept almost everything that I wrote [in dev-bot].'' (P8)
\end{quote}

Similarly, P2 described dedicating ``\textit{the first 15 minutes or so to provide a lot of rules}'' to the dev-bot, followed by testing the generated bot: ``\textit{I literally tested the bot at the end, unlike P1 who was simultaneously testing her bot.}'' (P2).
Subsequently, when the test-bot adhered to her rules, she expressed amazement: ``\textit{I was like, oh wow!}'' (P2).
This process instilled trust in the workflow, and she mentioned repeating it even when building the knowledge bot (e.g., P2 sent 13 utterances to the dev-bot before testing her knowledge bot).

\subsubsection{$(dev - test)^n$}
Two participants (P1 and P10), both with no prior programming experience, mentioned that they completely ``\textit{ignored representations}'' and solely interacted with the dev-bot and test-bot.
As P10 articulated: ``\textit{I think once I got a good hang of it, it was fast enough to write, test, write, test... that way it was pretty cool.}''
P1 did not click any of the representation components (tabs) even once, while P10 clicked them a total of three times.
It is important to note that using representation click data as a measure of engagement with the representation would be misleading.
For example, three participants (P5, P9, and P13) did not click on the representation tabs at all, yet they performed multiple direct representation edits to the default view logic component, confirming their engagement with the representation.

Despite the representations being presented in natural language, participants with no programming experience found it challenging to understand the meaning of terms like `logic', `variables', and `knowledge base'.
They also complained that the representations were written in a ``\textit{very formal language}'' which they found difficult to comprehend, and this demotivated them from reading or editing it.
For instance:
\begin{quote}
    ``I would not understand it (the representations) even if I went into the logic... Like here [in representation,] it says `score variable' which I don't understand, as in why it is a variable?... So I completely overlooked the center part... I do not know anything about coding. I was completely dependent on the [dev-bot] chat. If something was not working, I would just tell [the dev-bot].'' (P1)
\end{quote}
This workflow of testing the bot directly after an utterance to the dev-bot was quite prevalent, as each of our 14 participants engaged in it at least once, with an average of 4.9$\pm$3.3 times.
Of these, seven participants (four intermediate and three expert programmers) opted to test their bot after the initial utterance to the dev-bot in one of their projects.

\subsubsection{$rep_{edit}^n$ -- test}
In the log data, we found only 42 instances of direct edits to the representation, carried out by 12 participants.
As discussed in Section~\ref{sec:rep_usage}, participants performed representation edits for three primary reasons:
when their utterance to the dev-bot did not yield the outcome they were expecting,
when representation edits required less effort for a particular change compared to making the edit through utterances to the dev-bot (e.g., P11's update to the logic component: `\fontttc{For each continent, give \tcbtppink{20}\tcbtpgreen{5} questions to the user.}'), and
because representation edits were perceived to be faster in terms of execution time.

Let us delve into an example to illustrate a dev-bot utterance failure.
P6's utterance to the dev-bot was: `\fontttc{If the user is able to get the first question correctly, reward them with 100 points, after that if the user is able to get the next question right then reward them with 200 points and so on.}'
This resulted in the representation logic being updated as: `\fontttc{Increase score by 100 points for the first question, then increase by 200 points for each subsequent question.}', which didn't accurately capture P6's scoring scheme.
Hence, P6 made this direct edit to the logic representation: `\fontttc{If the user is able to get the first question correctly, reward them with 100 points, after that \tcbtppink{if the user is able to get the next question right then reward them with 200 points and so on.}\tcbtpgreen{increase the scores assigned by 100 with each subsequent question. So, if the user gets the second question correct, the user gets 200 score, 300 for third question, and so on}.}'
This, in turn, triggered an automatic update of the variables component with: `\fontttc{score: 0 \{Initial value: 0, Update rule: Increment by 100 * questionNumber after answering the questionNumber correctly.\}}'.
This example elucidates that direct representation edits were crucial for participants in refining the output bot's logic to align with their intended outcomes.
Participants highly valued the flexibility and control provided by the ability to directly edit the representations.

Participants also identified instances of `hallucination' errors in the representation, which are relatively common in LLM-based systems~\cite{hallucination-survey}.
For example, P7 did not define any knowledge base for her animal quiz, so she edited the logic accordingly: `\fontttc{1. When a new question is asked, randomly select a fun fact \tcbtppink{from the Animal Fun Facts section of KB}\tcbtpgreen{about Animals}.}'.
P14 made a very similar edit as well.
We found four instances of participants editing such hallucinations in the representations.
As expected, our \pbyc{} system, like any other AI system, has flaws.
Therefore, providing users with the means to rectify AI errors was highly appreciated.

Only three participants (2 expert and 1 intermediate programmer) performed bulk edits--i.e., three or more consecutive edits--using representations.
In their workflow, they began with an utterance to the dev-bot, but soon realized that they could achieve their goals ``\textit{faster}'' and ``\textit{more accurately}'' by directly editing different representation components. 
P13, for example, performed five consecutive representation edits, providing two of them here for context:
`\fontttc{Merit Certificate Link: initial value - None; update - based on degree, year, \tcbtppink{and}id\tcbtpgreen{, and semester}.}', and `\fontttc{Semesters: There are 8 semesters for any user, fall and spring semesters across 4 years. Current semester is summer \tcbtppink{of the current year}\tcbtpgreen{2023}.}'.
P13 started the chatbot development process by sending four messages to the dev-bot.
Once the output bot exhibited a satisfactory state, P13 subsequently proceeded with direct representation edits, followed by testing for validation.
It highlights the importance of empowering advanced users with flexible customization options, as it enabled them to fine-tune the bot's behavior more efficiently and precisely.

\subsection{Programming Experience using \pbyc{}}
The novel conversational programming approach offered by \pbyc{}, where a user could build a chatbot by conversing with another bot, was well-received by our participants.
They found it to be ``\textit{easy to use}'', ``\textit{easy to learn}'', ``\textit{fun}'', and has potential to democratize programming, 
Nonetheless, they also raised valid concerns regarding the applicability of such a tool in creating chatbots for intricate use cases and high-stakes settings.
Below we discuss these in detail.

\subsubsection{Conversational Programming}
Participants were in awe at the capability of \pbyc{} to generate a functional bot instantly just by natural language interaction with the dev-bot.
Additionally, they observed successful updates to (parts of the) previously defined representations.
For instance,
P7's initial utterance was `\fontttc{Create a bot on addition and multiplication of 2 digit numbers for visually impaired students. The text should be simple and understandable and not require sight to understand.}', which was subsequently followed by `\fontttc{Include one digit additions also}'. This resulted in the KB getting modified as: `\fontttc{This bot is designed to support the addition and multiplication of \tcbtpgreen{1-digit and} 2-digit numbers for visually impaired students.}'. E.g.,
\begin{quote}
    I could also go back on the prompts (utterances) that I made [to dev-bot], like first I asked it to do something, and then I could just go back and say no I don't want this, I want that. That was also picked fairly well... it's like changing just a part of a logic statement. It was doing that correctly. (P8)
\end{quote}
We found our participants readily updating the scoring mechanism and even the theme of their quiz/knowledge bots with minimal effort. 
This was particularly evident in the first task of quiz bot creation (refer Figure~\ref{fig:sample_rep} for an example), where participants enthusiastically added and updated several rules around scoring scheme, difficulty levels, availability of lifelines, provision of hints, and types of questions.
    




Despite their diverse levels of programming proficiency, all participants were impressed with \pbyc{}'s ability to effectively follow their (vague) instructions.
As P14 highlighted: ``\textit{Even though my instructions were vague, the chatbot somehow kind of went where I wanted it to go... it kind of already knew what a quiz is... so I mean, you know that that's kind of an advantage.}''.
This proficiency can be partially attributed not only to the robust GPT-4 model underpinning \pbyc{}, but also to the provision of initial (hidden) prompts within the quiz bot and knowledge bot templates, which facilitated an easy start for users in constructing their respective bots.
However, two participants lacking prior experience in prompting and programming expressed uncertainty in formulating their first utterance to the dev-bot:
\begin{quote}
    ``For the first message [to dev-bot], I was not sure how deep I should go, or at how much surface level it should be?... I was kind of scared in the beginning... I think it should have prompts to help me give prompts (utterances to dev-bot)... basically like `you can say this or say that', like suggestions.'' (P1)
\end{quote}
This uncertainty was likely due to the novelty of the task and the \pbyc{} platform, as P1 was a designer with no prior experience in programming, prompting, or chatbot development.
Notably, the majority of participants (12 out of 14) reported encountering a minimal learning curve in using \pbyc{}.
Echoing P1's suggestion, other participants recommended that the dev-bot should provide feedback, such as encouraging users to provide concise utterances to dev-bot, or guiding the document upload process for custom KB creation.
Additionally, P3 expressed uncertainty regarding the distinctions between the three representation tabs and proposed a method for seeking such clarifications from the dev-bot.
\begin{quote}
    ``Is there a functionality to ask questions [to dev-bot] itself?... When I asked, `what is KB?' [to dev-bot], it was like, `No update to KB'... Let's say if you're making a bot that's super complex, it would be helpful if let's say I forgot what I instructed initially, I could just ask it (dev-bot) to verify the bot's current behavior... Can I put a query asking, `hey have I put this in logic', or `what is the scoring mechanism'?... In that case, it (dev-bot) should just give me the answer rather than updating anything.'' (P3)
\end{quote}
The idea of employing the dev-bot not only for bot creation but also to assess the output bot's capabilities is intriguing.
We observed that participants with a higher level of programming proficiency gained insight into the current capabilities of the output bot by scrutinizing the representations. On the other hand, novice programmers may find it beneficial to rely on conversing with the dev-bot to assess its current understanding.
This bears resemblance to Whyline~\cite{whyline-chi04}, a debugging paradigm that enables programmers to ask `why' questions about their programs.

\subsubsection{Strategies}
While interacting with \pbyc{}, our participants devised several strategies, including transitioning from general to specific utterances and providing examples to dev-bot.
First, the majority of participants initiated their interaction with the dev-bot using a broad utterance, such as `\fontttc{Hello. I want to build a quizzing app that help learners of Punjabi language test their skills. You can start by asking questions on the meanings of basic words and phrases and keep on increasing the complexity as we move along.}' (P9), and `\fontttc{Develop a teaching assistant for an introductory HCI course in a university}' (P5).
They refined their instructions with specific details in subsequent utterances. For instance, `\fontttc{I notice that the answer to multiple choice questions is always the first option, can you make sure that the correct answer is always randomized?}' (P9) and `\fontttc{The students in this HCI course are beginners. Suggest them easy to use methodologies with concrete steps}' (P5), respectively.
P9 explicitly articulated this strategy:
``\textit{I started with a very general instruction }[to the dev-bot]\textit{, and then I kept on adding more specific things that I wanted to do.}''
This approach aligns with established practices in prompt engineering for LLMs~\cite{prompt-engineering-tech-azure, killer-prompt-landbot, genline-chi22}.

Second, in both the quiz bot and knowledge bot scenarios, our participants provided examples to guide the \pbyc{} system in developing their custom bot.
While most participants opted for quiz bots centered around popular themes, such as food or Bollywood, which did not necessitate an example question, P6 attempted a football quiz bot specific to particular awards. In this case, providing an example (`\fontttc{Add Multiple choice questions like: Q. Which of the following player won the Ballon d'or in 2022?}') proved instrumental.
P6 reflected on this experience, stating ``\textit{with just one question it was really able to generalize well}''.
This approach is similar to the concept of few-shot learning and is a conventional strategy in prompt engineering~\cite{prompt-engineering-tech-azure}.
Interestingly, we observed that participants with no prior programming and/or prompting experience were able to recognize and effectively employ this strategy.
For instance, P10, a novice programmer, adopted a similar approach when creating a quiz focused on Indian freedom fighters.

\subsubsection{Democratize Chatbot Development}
All participants unanimously agreed that \pbyc{} holds the potential to empower individuals, even those lacking prior programming experience, to effortlessly develop chatbots. P3 humorously remarked, ``\textit{This will take my job!}''.
P1, a designer, elaborated on the potentially transformative impact of \pbyc{} on her workflow:
\begin{quote}
    ``In my day-to-day life, I design something, and then wait for someone else to implement it. But here I was designing and developing both by myself... It will be very useful for me to test my ideas... If I am designing a bot, I can just quickly try it [using \pbyc{}] and see if it's actually working or not. That would be useful for me, because I won't have to go back to the developers and be like `oh, this is the flow', and they'll say, `this is not possible'. Then I have to come back and reiterate. I am heavily dependent on them... This (\pbyc{}) I think will make it way easier for me to just test what I've come up with.'' (P1)
\end{quote}
This further underscores the agency that \pbyc{} affords its users (discussed in Section~\ref{section:understanding}).
\pbyc{} not only demonstrated future potential, but in specific instances, due to the open-ended nature of the study task, participants created knowledge bots that could significantly assist in their daily routines. For example, P10, a novice programmer and IT manager of an organization, developed a bot to address IT-related inquiries, streamlining her responsibilities. She remarked:
\begin{quote}
    ``I did not have to think about what goes behind in terms of programming. All I had to do was give the instructions and it carried out perfectly... I only wish I could start using it (the IT Knowledge Bot I created) for all our IT operations. So if somebody comes to us, we have a ready-made answer... We can have answers for most things... [like] how do I get my SCR card?, where do I apply for GCR?'' (P10)
\end{quote}

The ease of access to chatbot development was not only embraced by novice and intermediate programmers, but even expert programmers acknowledged the value of \pbyc{} in expediting their initial progress and overall efficiency:
\begin{quote}
    ``When we start a new project, we have to put in the boilerplate code, as in the template codes and stuff that I see it (\pbyc{}) could do very well... With that, I could get started in just 5 minutes without worrying... without procrastinating.'' (P14)
\end{quote}
\begin{quote}
    ``If I had to code the same bot, it would have taken me like 2-3 hours, but this took 10-15 minutes'' (P8)
\end{quote}

\subsubsection{Missing Testing and IDE}
Despite recognizing and valuing the potential of \pbyc{} in chatbot development, our participants expressed reservations about its suitability, particularly for creating high-stakes, extensively-tested, real-world chatbots.
They delineated several underlying concerns.
Foremost among these was the apprehension regarding the applicability and effective use of representations, particularly when dealing with a large number of intricate rules present in real-world bots,
as P13 illustrated: ``\textit{When the number of rules in the logic is very high... then I wouldn't be able to review all of them... I wouldn't be confident it }(\pbyc{})\textit{ has correctly generated the logic.}''.
Furthermore, participants were unsure about deploying a chatbot generated by a LLM in high-stakes scenarios due to the stochastic nature of LLMs:
\begin{quote}
    ``In my quiz I asked it to give four options. Sometimes it presented the options as 1,2,3,4 and sometimes as A,B,C,D... Important point is that if I didn't test it enough, I wouldn't have known this error, right? 
    This thing I spotted, so I fixed it... but there might be many more things which I didn't spot.'' (P14)
\end{quote}

Conducting exhaustive manual testing to cover all possible cases was deemed ``\textit{impossible}'' and impractical.
Hence, participants advocated for a mechanism to write unit tests for rigorous testing of the generated bot:
\begin{quote}
    ``I still think that there's no way for me to understand if it has `completely' understood what I am saying... I think there should be some way for the developer to do some rigorous testing. I think the way I interacted with it was... that I started building something basic, I tested it, I added more instructions, and then I tested it. So I added instructions in an increasing complexity... it would be good if I could also write some unit tests on top of it in increasing complexity... That will give me some assurance.'' (P5)
\end{quote}
Ultimately, our participants, particularly the expert programmers, wanted more capabilities of Integrated Development Environments (IDEs) such as generation and management of unit tests, to have higher quality assurance.
They also felt that such quality assurance measures would be a requirement for real-word chatbot development tasks.

%% file: tables/results.tex
\begin{table}[]
\centering
\caption{Descriptive statistics from the log data}
\label{table:results}
\resizebox{\textwidth}{!}{%
\begin{tabular}{r|cccc|cccc|cccc}
\hline
\multicolumn{1}{c|}{} &
  \multicolumn{4}{c|}{\textbf{Quiz Bot}} &
  \multicolumn{4}{c|}{\textbf{Knowledge Bot}} &
  \multicolumn{4}{c}{\textbf{Overall}} \\ \cline{2-13} 
\multicolumn{1}{c|}{} &
  \textbf{Count} &
  \textbf{/user} &
  \textbf{/bot} &
  \textbf{/msg} &
  \textbf{Count} &
  \textbf{/user} &
  \textbf{/bot} &
  \textbf{/msg} &
  \textbf{Count} &
  \textbf{/user} &
  \textbf{/bot} &
  \textbf{/msg} \\ \hline
\textbf{Projects} &
  17 &
  \begin{tabular}[c]{@{}c@{}}m=1.3\\ sd=0.5\end{tabular} &
  - &
  - &
  16 &
  \begin{tabular}[c]{@{}c@{}}1.1\\ 0.4\end{tabular} &
  - &
  - &
  33 &
  \begin{tabular}[c]{@{}c@{}}2.4\\ 0.5\end{tabular} &
  - &
  - \\ \hline
\textbf{dev-bot msg} &
  132 &
  \begin{tabular}[c]{@{}c@{}}10.2\\ 3.7\end{tabular} &
  \begin{tabular}[c]{@{}c@{}}7.8\\ 4.4\end{tabular} &
  - &
  82 &
  \begin{tabular}[c]{@{}c@{}}6.3\\ 4.6\end{tabular} &
  \begin{tabular}[c]{@{}c@{}}5.5\\ 3.0\end{tabular} &
  - &
  214 &
  \begin{tabular}[c]{@{}c@{}}15.3\\ 6.9\end{tabular} &
  \begin{tabular}[c]{@{}c@{}}6.7\\ 3.9\end{tabular} &
  - \\ \hline
\textbf{dev-bot length} &
  2885 &
  \begin{tabular}[c]{@{}c@{}}221.9\\ 104.2\end{tabular} &
  \begin{tabular}[c]{@{}c@{}}169.7\\ 113.2\end{tabular} &
  \begin{tabular}[c]{@{}c@{}}21.9\\ 16.4\end{tabular} &
  1635 &
  \begin{tabular}[c]{@{}c@{}}125.8\\ 100.4\end{tabular} &
  \begin{tabular}[c]{@{}c@{}}109.0\\ 74.7\end{tabular} &
  \begin{tabular}[c]{@{}c@{}}19.9\\ 15.5\end{tabular} &
  4520 &
  \begin{tabular}[c]{@{}c@{}}322.8\\ 161.2\end{tabular} &
  \begin{tabular}[c]{@{}c@{}}141.25\\ 100.4\end{tabular} &
  \begin{tabular}[c]{@{}c@{}}21.1\\ 16.0\end{tabular} \\ \hline
\textbf{test-bot msg} &
  248 &
  \begin{tabular}[c]{@{}c@{}}19.1\\ 8.5\end{tabular} &
  \begin{tabular}[c]{@{}c@{}}15.5\\ 9.3\end{tabular} &
  - &
  176 &
  \begin{tabular}[c]{@{}c@{}}12.6\\ 7.6\end{tabular} &
  \begin{tabular}[c]{@{}c@{}}11.7\\ 6.2\end{tabular} &
  - &
  424 &
  \begin{tabular}[c]{@{}c@{}}30.3\\ 8.9\end{tabular} &
  \begin{tabular}[c]{@{}c@{}}13.7\\ 8.0\end{tabular} &
  - \\ \hline
\textbf{test-bot length} &
  557 &
  \begin{tabular}[c]{@{}c@{}}42.8\\ 28.2\end{tabular} &
  \begin{tabular}[c]{@{}c@{}}34.8\\ 28.4\end{tabular} &
  \begin{tabular}[c]{@{}c@{}}2.2\\ 1.9\end{tabular} &
  1010 &
  \begin{tabular}[c]{@{}c@{}}72.1\\ 62.4\end{tabular} &
  \begin{tabular}[c]{@{}c@{}}67.3\\ 62.5\end{tabular} &
  \begin{tabular}[c]{@{}c@{}}5.7\\ 5.9\end{tabular} &
  1567 &
  \begin{tabular}[c]{@{}c@{}}111.9\\ 70.8\end{tabular} &
  \begin{tabular}[c]{@{}c@{}}50.5\\ 50.0\end{tabular} &
  \begin{tabular}[c]{@{}c@{}}3.7\\ 4.5\end{tabular} \\ \hline
\textbf{Rep edit} &
  19 &
  \begin{tabular}[c]{@{}c@{}}1.7\\ 0.8\end{tabular} &
  \begin{tabular}[c]{@{}c@{}}1.7\\ 0.8\end{tabular} &
  - &
  23 &
  \begin{tabular}[c]{@{}c@{}}3.0\\ 0.7\end{tabular} &
  \begin{tabular}[c]{@{}c@{}}3.0\\ 0.7\end{tabular} &
  - &
  42 &
  \begin{tabular}[c]{@{}c@{}}3.0\\ 1.7\end{tabular} &
  \begin{tabular}[c]{@{}c@{}}2.25\\ 0.9\end{tabular} &
  - \\ \hline
\textbf{Rep click} &
  107 &
  \begin{tabular}[c]{@{}c@{}}21.4\\ 15.5\end{tabular} &
  \begin{tabular}[c]{@{}c@{}}17.8\\ 15.6\end{tabular} &
  - &
  98 &
  \begin{tabular}[c]{@{}c@{}}19.6\\ 12.7\end{tabular} &
  \begin{tabular}[c]{@{}c@{}}16.3\\ 8.1\end{tabular} &
  - &
  205 &
  \begin{tabular}[c]{@{}c@{}}34.2\\ 13.6\end{tabular} &
  \begin{tabular}[c]{@{}c@{}}17.1\\ 11.9\end{tabular} &
  - \\ \hline
\end{tabular}%
}
\end{table}

%% file: content/6_discussion.tex
\section{Discussion}
Our work presents a qualitative and quantitative evaluation of an LLM-powered system that facilitates chatbot development using natural language interaction and representations.
While our primary focus is on addressing the understanding mismatch problem between users and the system through representations, our findings are intertwined with the programming expertise of our participants.
Below we discuss that in detail and also provide insights into the usability, trustworthiness, and learnability of LLM-based systems.

\subsection{Novice vs Intermediate vs Expert Programmers}
Similar to recent works \cite{abstractiongap-chi23, programmer-assistant-iui23}, in our study, we recruited participants with varying proficiency levels in programming to assess the broader applicability of our proposed \pbyc{} methodology, which in turn enriched the depth and diversity of our findings.
One of our key findings is the novel mode of interaction mechanism termed `glancing' at the representations.
The extent of glancing interaction exhibited a direct correlation with the participant's programming expertise.
Expert programmers engaged in comprehensive reading, re-reading, and making substantial edits to the representations.
Intermediate programmers primarily glanced to validate the system's interpretation and performed minimal edits, limited to deletions and minor updates, without additions. 
In contrast, novice programmers engaged with the representations at a minimal level, often choosing to ignore them entirely.
A longitudinal study would provide insight into user's progression in utilizing representations as they naturally transition from novice to intermediate, and from intermediate to expert programmers.
As prior research~\cite{novice-psychology-survey81} indicates that novice programmers learn by articulating technical details in their own words.
Such a longitudinal investigation of \pbyc{} usage would shed light on the role of LLM-based systems in advancing programming proficiency, as the representations contribute to increased system's transparency. 

Apart from interaction patterns, notable differences in adopted and preferred workflows were observed across varying levels of programming expertise. 
Novice programmers primarily relied on interacting with the dev-bot, followed by testing the generated test-bot, often conducting tests subsequent to each utterance to the dev-bot.
In contrast, intermediate and expert programmers utilized representations to assess the system's interpretation of their input utterance to the dev-bot, instead of testing the output bot after each utterance.
Furthermore, a few intermediate and expert programmers recognized direct edits to the representation as a more effective method of updating, thus employing a nuanced combination of interactions with the dev-bot, scrutinizing the updated representation for validation, directly editing the representation, and, upon satisfaction, engaging with the test-bot to evaluate for edge cases and other unidentified issues.
The flexibility afforded by the \pbyc{} system proved instrumental in enabling participants to explore these diverse workflows and select the ones that aligned with their specific requirements for a particular scenario.
Moreover, it conferred upon them a sense of agency and control, dimensions that have been reported as lacking in prior LLM-based systems~\cite{usability-aiprog-icse24, explainability-iui22, hbr-prompt-notfuture, johnny-cant-prompt-chi23}.
Future LLM-powered systems should aim to offer such flexible customization options, to not only enhance user's efficiency and productivity, but also to identify and rectify LLM errors.

Finally, our participants' identification of potential use cases for \pbyc{} was also significantly influenced by their respective levels of programming proficiency.
Novice programmers preferred employing \pbyc{} for developing end-to-end applications, while intermediate programmers cited its utility in rapid prototyping.
In contrast, expert programmers expressed a desire to use it for generating boilerplate code. 
Nevertheless, these expert programmers needed a way to systematically test the output bot (similar to \cite{johnny-cant-prompt-chi23}), and were skeptical of using an LLM-powered system for developing high-stakes end-user products.
This hesitancy may be attributed to the fact that source code remained concealed in \pbyc{} from the user's view.
Additionally, there was a prevailing perception that the system was primarily suited for developing no-code chatbots.
In future iterations, we intend to extend the applicability of \pbyc{} methodology to web application development, leveraging a blend of JavaScript, HTML, and CSS, wherein showing the source code as a distinct component within the representation, alongside logic, variables, and knowledge base, would add value.
In this scenario, a distinct template will be provided to facilitate web application development, the dev-bot pane will retain its current format, and the test-bot will be substituted with the outcome generated by the web application.
Likewise, we aim to extend support for use cases involving Python-based data analysis (similar to \cite{abstractiongap-chi23}).
Employing the same (\pbyc{}) methodology for exploring multiple use cases is an idea worth-exploring in the LLM space.

It is important to note that in all the aforementioned aspects--interaction patterns, workflows, and identified use cases--no rigid demarcation was observed among users of different programming expertise. Nonetheless, we present here the sentiments expressed by the respective user groups, which may not necessarily be representative of all users within their respective categories.

\subsection{Usability, Trustworthiness, and Learnability}
While the majority of participants found the \pbyc{} system to be user-friendly, requiring minimal learning effort, and trusted it for chatbot development, we have identified areas for potential improvement in these critical metrics.
Most participants reported that representations aided their understanding of the system's interpretation of their utterances and found direct edits to representations effective.
However, novice programmers faced challenges in this regard.
To enhance the usability of representations for novice programmers, first, it is advisable to refrain from employing technical jargon, such as `knowledge base' and `variables', instead, terminology that is more relatable to non-programmers, such as `information library' and `data holders' could be used.
Second, it is plausible that the three defined components may not offer the most optimal representation.
There is room for exploring alternative representations.
Participants suggested a flow-diagram-based hierarchical representation as a potential improvement over our linear representations.
In future iterations, experimenting with different components could contribute to enhancing the usability and consumption of representations.
Third, the ambiguity in distinguishing between KB, logic, and variables, could be resolved through more extensive training. 
In the current study, users were encouraged to explore the system with minimal guidance, hence training was limited to a concise 2-minute instructional video.
Finally, in addition to utilizing the dev-bot for receiving instructions on bot construction, incorporating a mechanism for users to acquire more comprehensive insights into the \pbyc{} system, as well as details about the specific chatbot they are developing, by interacting with the dev-bot, would augment the overall value delivered to users.

With respect to trust, users expressed confidence in our system's ability to rapidly generate prototypes for testing their ideas and produce boilerplate code. However, they hesitated to trust it for developing chatbot products intended for external deployment, primarily due to the absence of a systematic testing infrastructure within the tool.
The inherent stochasticity in the output of the LLM also significantly impacted trust, as the bot did not consistently adhere to the rules outlined in the representation in each iteration of the test-bot.
To mitigate this, we regulated the randomness and creativity of the generated text in GPT-4 by setting the `temperature' parameter to a value of 0.3. While this yielded a more consistent output, further fine-tuning of the temperature parameter is warranted.
Moreover, the occurrence of hallucinations~\cite{hallucination-survey}, a recognized issue with LLMs, contributed to diminished trust.
Participants noted that access to representations afforded them a means to identify and rectify these hallucination instances.

Finally, while most users quickly learned how to use our system and efficiently crafted complex bots in a short time frame, a subset encountered the `Blank Page Syndrome'~\cite{blank-page-syndrome}, a phenomenon wherein users, faced with a blank canvas, experience uncertainty about where to commence.
To address this challenge, we propose a paradigm shift from a user-driven interaction with dev-bot to a bot-driven conversation.
In this revised system, based on the project description provided by the user during chatbot creation from a given template, the dev-bot will pose pertinent questions, thus guiding the user through the chatbot building process.
This approach holds the potential to enhance the system's usability and expedite the learning process.

%% file: content/7_conclusion.tex
\subsection{Limitations}
We acknowledge several limitations of this work.
First, our participant pool leaned towards professionals in STEM-related domains.
It is important to note that our sample is not a representative cross-section of the broader population. 
Consequently, we refrain from asserting or inferring the prevalence of specific behaviors or beliefs within the general populace. 
Instead, our focus is on delineating behavioral patterns within a demographic that we anticipate possesses a disproportionately higher likelihood of early adoption of LLM-based tools.
Second, the small sample size limited our analyses. A larger number of participants is required to identify broader trends. 
However, similar studies in the past have had comparable sample sizes, e.g., \citet{johnny-cant-prompt-chi23}'s study involved 10 participants, and \citet{abstractiongap-chi23}'s study included 24 participants.

\section{Conclusion}
To address the gap between the user's perception of the AI system's understanding and the system's actual understanding, we introduced the \pbyc{} methodology, utilizing representations to convey the LLM's comprehension. 
Through a comprehensive study with participants of varying programming expertise levels, we found that representations not only enhances user understanding, agency, and trust, but also streamlines the development process, making it more transparent and efficient.
The synergy between natural language-based development with LLMs and representations offers a promising avenue for revolutionizing software development, making it more accessible and user-friendly for a broader audience.